\documentstyle[aps,prl,twocolumn,epsf]{revtex}
\begin{document}
\draft

\title{Irreversibility Line in Nb/CuMn Multilayers with a Regular Array of Antidots}
\author{C. Attanasio, T. Di Luccio, L.V. Mercaldo, S.L.
Prischepa,
R. Russo,
M. Salvato, and L. Maritato}
\address{Dipartimento di Fisica and INFM, Universit\'a degli Studi di Salerno,
Baronissi (Sa), I--84081, Italy}
\author{S. Barbanera}
\address{IESS-CNR, Via Cineto Romano 42, Roma, I--00100, Italy}
\author{A. Tuissi}
\address{CNR TeMPE, Via Cozzi 53, Milano, I--20125, Italy}

\date{\today}
\maketitle

\begin{abstract}
The transport properties of Nb/CuMn multilayers with a regular
array of electron beam lithography obtained antidots have been
measured at different temperatures in the presence of external
perpendicular magnetic fields. Hysteretical $I-V$ characteristics
have been observed which disappear when approaching the upper
critical magnetic field curve $H_{c2}(T)$. Comparing these data
with other results (Arrhenius plots of resistive transition
curves, logV-logI characteristics) we have been able to relate the
onset of the hysteresis to the presence of an irreversibility
line. We discuss several possible mechanisms to clarify the nature
of this line. Among them the most plausible seems to be the vortex
melting mainly induced by quantum fluctuations.
\end{abstract}


\section{Introduction}

The task of increasing the critical current density $J_c$ in
superconducting materials has always been a widely studied subject
\cite{Camp}, gaining even more interest after the discovery of
high temperature superconductors (HTS) \cite{Blatt1}. These
studies are strictly related to the knowledge of the flux line
pinning mechanism and to the reduction of the vortex mobility.
Introducing  artificial defects in superconducting materials as,
for example, non superconducting distributed phases
\cite{Nb3Sn,Shu}, columnar tracks of amorphous material obtained
by high energy ions \cite{Civ} or geometrical constrictions such
as channels or dots \cite{Pru,Crabtree}, is a very useful tool for
a better understanding of the vortex dynamics and for obtaining
higher $J_c$ values. The recent developments of the submicrometer
electron beam lithographic techniques have made possible to reduce
the typical size of these geometrical constrictions to values much
smaller than the period of the vortex lattice, and comparable to
those of typical superconducting coherence lengths
\cite{Shu,Bae,Ket,Word}. Many experiments, performed on systems
with such a regular array of defects\cite{Bae,Ket,Word,Fio,Lykov}
and several numerical simulations \cite{Nori1,Nori2}, have been
focused on the study of the vortex properties at low magnetic
fields close to the matching fields $H_n=n_p \Phi_0$ (here $n_p$
is the pins concentration and $\Phi_0$ is the flux quantum).

Other studies, performed at higher magnetic fields, have been related to the
analysis of the vortex lattice shear stress in superconducting layered systems
with the presence of artificially obtained weak pinning channels embedded in a
strong pinning environment \cite{Pru,Crabtree,Mart,Past}.
Due to the possibility of varying in a controlled way many different parameters
 such as the density, the dimensionality and the nature of the pinning centers
 \cite{Shu,Ket,Philmag}, the study of the vortex dynamics in artificially layered
 conventional superconducors with the simultaneous presence of a regular array of
 pinning centers perpendicular to the layers is of great interest. Moreover, the
 natural layered structure of HTS allows to use these artificial systems as a model
 to help in discriminating among intrinsic and dimensional effects in the transport
 properties of HTS compounds \cite{Multasmodel}.

In this paper we report on current-voltage ($I-V$) characteristic
and resistance versus temperature, $R(T,H)$, measurements, in
perpendicular external magnetic fields $H$, performed on two
different series of Nb/CuMn multilayers with a square array of
antidots. The choice of a superconducting (Nb)/spin glass (CuMn)
layered system could be interesting particularly in view of the
use as a model of HTS compounds. The two series have antidots with
the same diameter, $D \approx 1 $ $\mu m$, and different lattice
distances between the antidots: $d \approx 2$ $\mu m$ in one
series and $d \approx 1.6$ $\mu m$ in the other. The experiments
have been performed on different samples having different
anisotropies for each series. The $I-V$ curves measured at
different temperatures and at different values of $H$ have shown,
in regions of the $H-T$ phase diagram depending on the anisotropy
of the system, a hysteretic behavior with sudden voltage jumps,
which disappers approaching the critical field $H_{c2}(T)$ curve.
From the analysis of the curvature of the logarithmic $I-V$
characteristics and from the study of the shape of the Arrhennius
plots of the resistive transition curves, we have been able to
relate the disappearing of such a hysteretical behavior to the
presence of an irreversibility line (IL). We have discussed
different possible mechanisms responsible for this IL. Among them
the most plausible for our samples seems to be the vortex melting
mainly induced by quantum fluctuations \cite{Blatt2}.

\section{Experimental}

Regular arrays of antidots have been fabricated using Electron
Beam Lithography to pattern the resist on a 2 inch Si (100) wafer.
The Nb and CuMn layers were deposited by a dual-source
magnetically enhanced dc triode sputtering system with a movable
substrate holder onto the patterned resist and the final
structures were obtained by the lift-off technique. Single
antidots have a circular geometry and the antidot array is
arranged in a square lattice configuration, see figure \ref{fig1}.
The total area covered by the array is a square of 200 $\times$
200 $\mu m^2$ with four separate contact pads connected to the
vertices. Eight replicas of this structure are present on the same
Si wafer to allow the fabrication of a series of multilayered
samples with the same Nb thickness and variable CuMn thicknesses
in only one deposition run \cite{Maritato}. The resist used is UV
III from Shipley; UV III is a chemically amplified photoresist for
the deep-UV range, but is widely used as an e-beam resist because
it provides a good tradeoff between reasonable sensitivity and
high resolution. In our case it has been used as a positive
resist, that is, the resist is retained where unexposed. A resist
film thickness of 5800 \AA\ has been obtained spinning the wafer
at 1800 rpm. The resist has been exposed using a Leica Cambridge
EBMF 10 system operated at 50 kV. Several tests have been carried
out in order to achieve structure profiles most suitable for
lift-off. In particular the desired profile is that showing a
moderate undercut. Also, developing times and post-exposure
treatment have been optimized for profile improvement. After
developing and post-baking at 130 $^\circ$C the samples underwent
RIE in $O_2$ for 30 $s$ at 25 W rf power and 14 Pa oxygen
pressure, in order to completely remove residual resist in the
exposed areas. This treatment lowered the resist thickness to
about 5000 \AA. Fig.1 shows a Scanning Electron Microscope (SEM)
image of the typical result of the fabrication process: in this
case the antidots nominal diameter is 1 $\mu m$ and the nominal
period of the structure is 2 $\mu m$.

\vspace{.5cm}
\begin{figure*}[h]
\begin{center}
\leavevmode \epsfxsize=7cm \epsfysize=4cm \epsffile{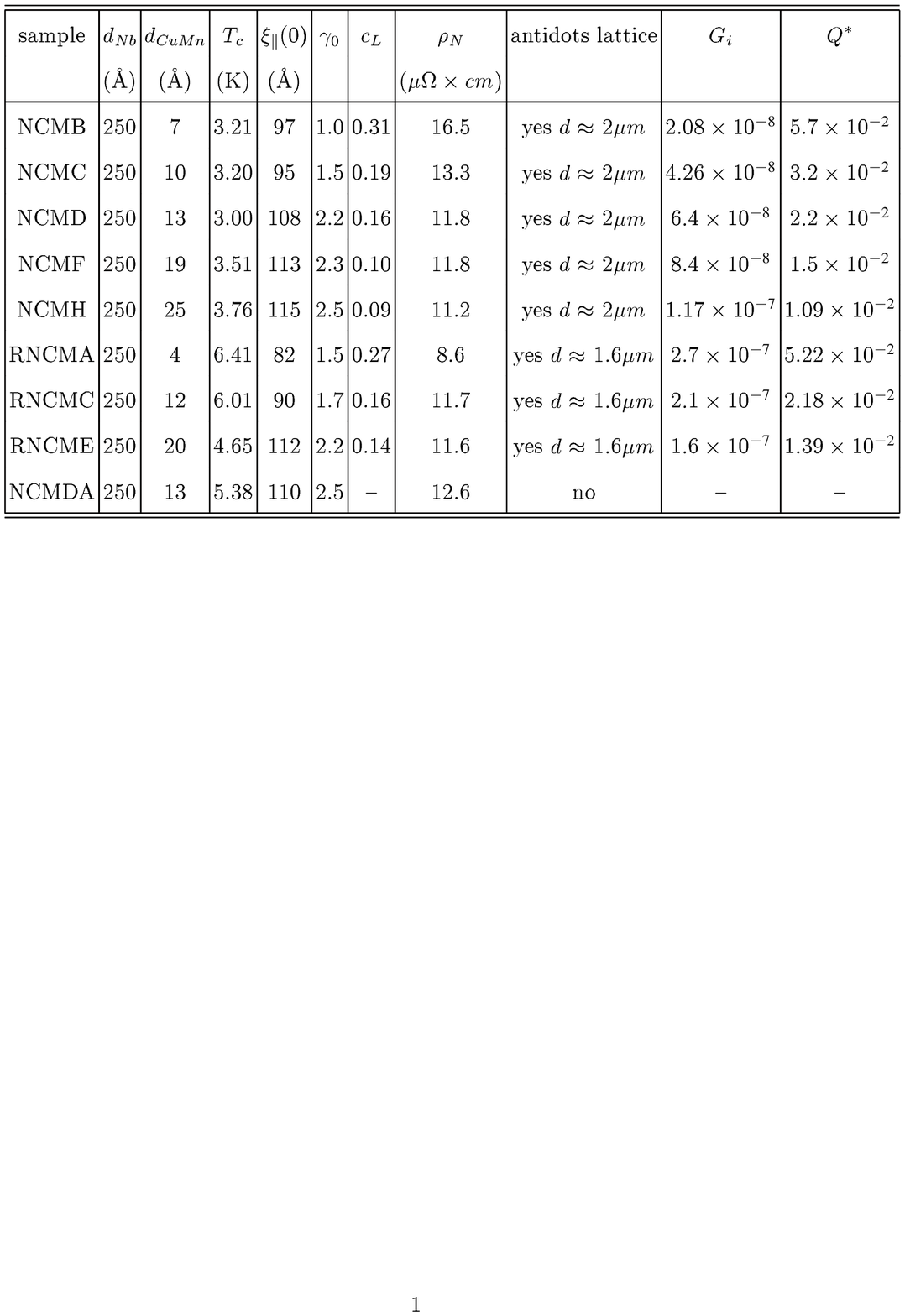}
\end{center}
\end{figure*}
Table I. Some of the relevant sample features. $\rho_N $ is the
resistivity at $T$=10 K. See the text for the meaning of the other
quantities.

\vspace{0.5cm}

The number of bilayers of Nb and CuMn is always equal to six. The
first layer is CuMn, the last one is Nb. The Nb nominal thickness,
$d_{Nb}$, is 250 \AA\ for all the samples in both the series. The
CuMn thickness, $d_{CuMn}$, has been varied from 7 \AA\ to 25 \AA\
in one series ($d \approx 2 \mu$m) and from 4 \AA\ to 20 \AA\ in
the other ($d \approx 1.6 \mu$m). The Mn percentage is always
equal to 2.7. For reference a Nb/CuMn multilayer with
$d_{CuMn}$=13 \AA\ (Mn \%=2.7) without antidots, but with the same
configuration (200$\times$200 $\mu m^2$ square and four pads
connected to the vertices), has been fabricated. The samples
parameters are summarized in Table I. $I-V$ characteristics have
been registred at $T \le 4.2$ K using a dc pulse technique. The
temperature stabilization, during the acquisition of curves in the
helium bath, was better than $10^{-2}$ K. The magnetic field was
obtained by a superconducting Nb-Ti solenoid. From the measured
temperature dependencies of the perpendicular and parallel upper
critical field $H_{c2}$ (obtained at half of the resistive
transitions $R(T,H)$) we deduced the values for the parallel and
the perpendicular coherence length at zero temperature,
$\xi_\parallel(0)$ and $\xi_\perp(0)$ respectively, and then the
value of the anisotropic Ginzburg-Landau mass ratio
$\gamma_0=\xi_\perp (0) /\xi_\parallel (0)$ \cite{Tin}. The
different values of the critical temperatures for the samples of
the two series having similar values of $d_{Nb}$ and $d_{CuMn}$
are probably related to the different Nb quality obtained in the
two deposition runs.

Figure \ref{fig2}a shows $I-V$ curves for the sample NCMF at
$T$=2.60 K for different applied magnetic fields in the range
0.03$<H<$0.70 Tesla. At low magnetic fields the curves present
hysteresis, not due to thermal effects, when registered both in
forward and backward directions, i.e. increasing and decreasing
the current \cite{Philmag}. Such hysteresis becomes smaller when
$H$ is increased disappearing completely, in the limit of our
experimental accuracy, when approaching $H_{c2}$. These features
have been repeatedly obtained for the same sample in different
measurements and are typical for all the samples of the two series
with antidots. In figure \ref{fig2}b the $I-V$ characteristics for
the sample NCMD are plotted in the temperature range 2.3
K$<T<$2.91 K for $\mu_0H$= 0.03 T. In this case the hysteresis
disappears approaching $T_c$. The $I-V$ curves for the sample
NCMDA without antidots are always smooth and parabolic-like,
typical of a type II superconductor.

In figure \ref{fig3}, we plot the $I-V$ curves in double
logarithmic scale for the sample RNCMC at $T$=2.96 K for different
values of the magnetic field. A change in the curvature of the
logV-logI curves clearly occurs at low voltages; this is usually
related to the presence of an IL in the $H-T$ phase diagram
\cite{Berg}, given by the points at which the logV-logI curve is
linear. In all the samples measured, these points are always very
close to the points at which the hysteresis in the $I-V$ curves
disappears. As an example, in the figures \ref{fig4}(a,b,c) are
shown the $H-T$ phase diagrams for the samples NCMB, NCMC and
NCMH, respectively. Circles distinguish the two dif-

\begin{figure}[h]
\begin{center}
\leavevmode \epsfxsize=7cm \epsfysize=6cm \epsffile{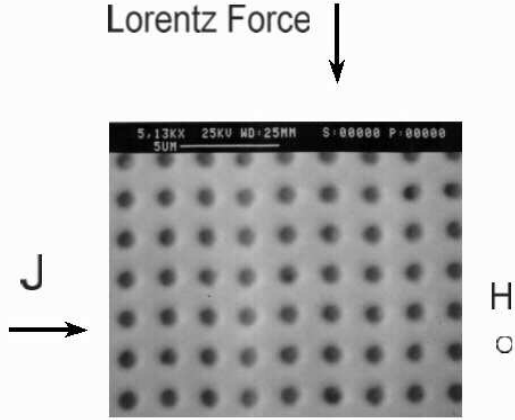}
\end{center}
\caption{\label{fig1}{Scanning Electron Microscope picture of a
Nb/CuMn sample with a square lattice of antidots ($D \approx 1$
$\mu m$ and $d \approx 2$ $\mu m$).}}
\end{figure}

\vspace{1.cm}
\begin{figure}[h]
\begin{center}
\leavevmode \epsfxsize=8cm \epsfysize=5cm \epsffile{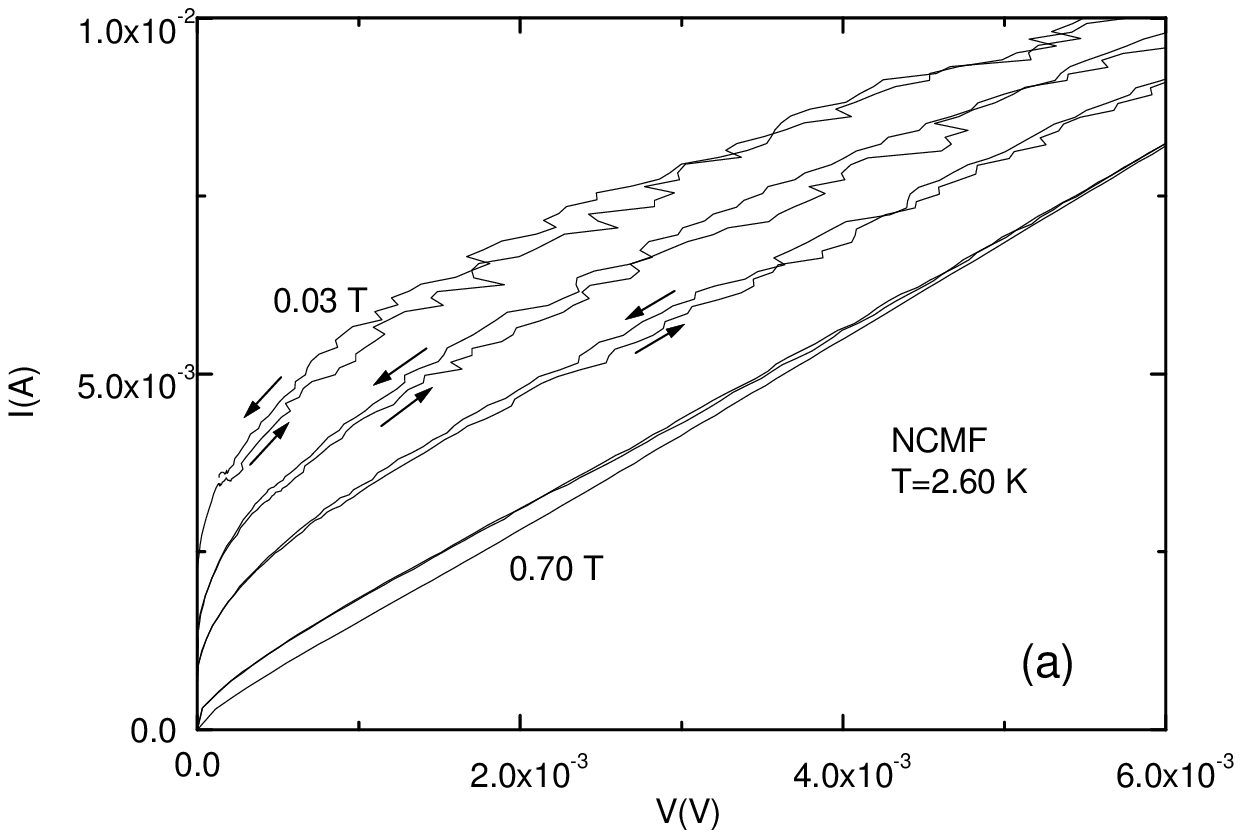}
\end{center}
\end{figure}

\vspace{.9cm}
\begin{figure}[h]
\begin{center} \leavevmode \epsfxsize=8cm \epsfysize=5cm
\epsffile{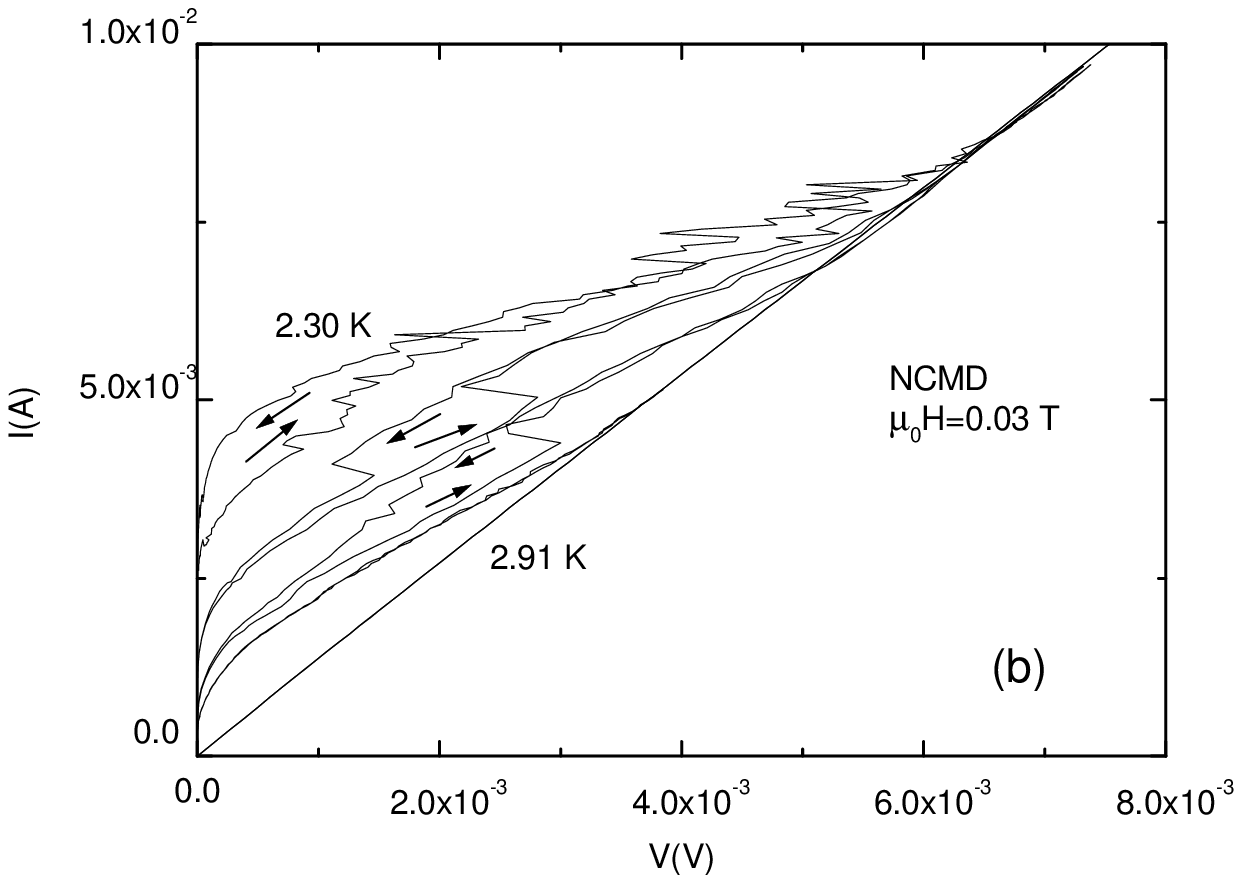}
\end{center}
\caption{\label{fig2}{(a) Current-Voltage characteristics for the
multilayer NCMF at $T=2.60$ K for different magnetic field values.
The increasing values are: 0.03; 0.06; 0.15; 0.55; 0.70 T. (b)
Current-Voltage characteristics for the multilayer NCMD at
$H=0.03$ T for different temperatures. The increasing values are:
2.30; 2.34; 2.60; 2.80; 2.91 K.}}
\end{figure}

\vspace{1cm}
\begin{figure}[h]
\begin{center}
\leavevmode \epsfxsize=7.5cm \epsfysize=5cm \epsffile{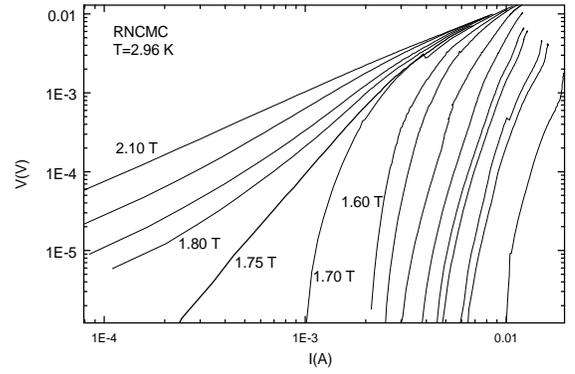}
\end{center}
\caption{\label{fig3}{Logarithmic Current-Voltage characteristics
for the sample RNCMC at $T=2.96$ K for magnetic fields, increasing
counterclockwise, of 0.20; 0.40; 0.50; 0.80; 0.90; 1.10; 1.30;
1.50; 1.60; 1.70; 1.75; 1.80; 1.85; 1.95; 2.10 T. The numbers
indicate some values of the magnetic field. The melting field is
estimated to be equal to 1.75 T.}}
\end{figure}

\noindent ferent regimes of the vortex dynamics as determined from
the disappearance of the hysteresis in the $I-V$ curves; up
triangles indicate points where change in the curvature of the
logV-logI occurs; squares correspond to the values of the
perpendicular critical magnetic field. It is then evident that the
change in the hysteretical behavior of the $I-V$ characteristics,
is related to the IL as defined by the change in the logV-logI
curvature. It is also interesting to note that the position of the
IL in the $H-T$ plane moves away from the $H_{c2}$ line as the
anisotropy of the samples increases.

Another way to determine the presence of an IL in the $H-T$ phase
diagram of a superconductor is the study of the Arrhenius plot of
the resistance versus the temperature curves \cite{Multasmodel}.
In figure \ref{fig5} the Arrhenius plot of the transiton curves,
recorded using a bias current of 2 mA, of the sample RNCMA is
shown at different perpendicular magnetic fields. A well defined
field dependent temperature $T^*$ separates two zones with very
different activation energies. In particular, at $T<T^*$ a sudden
increase in the Arrhnius slope signals a transition in the
transport properties of the sample which can be related to the
presence of the IL. In figure \ref{fig6} it is shown the measured
$H-T$ phase diagram  for the sample RNCMC. The solid squares
correspond to the perpendicular magnetic fields; the open squares
correspond to the points in the $H-T$ plane at which the onset of
hysteresis in the $I-V$ curves takes place, the open diamonds are
defined taking into account the change of curvature of the
logI-logV curves, and the open circles and up triangles are the
points at which the slope in the Arrhenius plot of the transition
curves, taken respectively at a bias currents of 2 mA and 6 mA,
change as viewed in figure \ref{fig5}. It is evident that also in
this case all the points (open symbols) fall again on the same
curve, which can be identified as an IL.

\section{Discussion}

In all the measurements performed the bias current was applied in
the plane of the film and perpendicular to the direction of the
external magnetic field, see figure \ref{fig1}. In this
configuration, the Lorentz force acting on the vortices, tends to
move them along the channels in between the antidots. On the other
hand, in the narrower zones between adjacent antidots, the current
density is much higher than in the channels, so that we locally
have weaker pinning centers in these parts of the sample.
Therefore, we can look at our superconducting layered system as
made of alternating zones of strong pinning (the channels along
the direction of action of the Lorentz force in figure
\ref{fig1}), and weak pinning (the narrower zones between adjacent
antidots), similarly to other cases reported in the literature
\cite{Pru,Crabtree,Mart,Past}. The value of the matching field in
both the series is very small, $H_n \approx 5$ Oe and also $d \gg
\xi(T)$, where $\xi(T)$ is the temperature dependent coherence
length which in our multilayers has typical values of about 100
\AA. Therefore, in the superconducting state it is possible to
have a vortex lattice inside the channels between the antidots.
The pinning of these interstitial vortices is determined mostly by
the intrinsic properties of the superconducting materials
\cite{Word,Bae1}.

To start the analysis of our data we have first to determine the
dimensionality of the vortex lattice in our samples. At magnetic
fields lower than a critical value $H_{cr} \approx 4 \Phi_0 /
\gamma_0^2 s^2$, where $s$ is the interlayer distance between
superconducting layers, the vortices in adjacent layers are
strongly coupled and vortex lines behave as three domensional
(3D). For our samples $H_{cr} \approx 10^3$ T, well above any
applied fields, and we can exclude the appearance of decoupling of
the vortex lines in our layered systems \cite{Vin}.

On the other hand, when the shear modulus $\rm c_{66}$ of a vortex
lattice is much lower than the tilt modulus $ \rm c_{44}$, the
system can behave bidimensionally. In principle, thermal
fluctuations could cause tilt deformations in the vortex line. The
critical thickness value $d_{cr}$ of a film beyond which thermal
fluctuation induced tilt deformations become relevant is given by
\cite{Glaz}

\begin{equation}
d_{cr} \approx {4.4 \xi_{\parallel} \over \sqrt{h(1-h)}} \label{eq:dcr}
\end{equation}

\noindent where $h=H/H_{c2}$. In multilayers, however, the
$d_{cr}$ value is reduced by a factor $\gamma_0^2$ \cite{Glaz}
which for our samples gives, in the range of measured temperatures
and magnetic fields, $d_{cr}^{multi} \approx 200 \div 300$ \AA.
Typical thickness for our samples are in the range $1500 \div
1700$ \AA. This means that the vortex lattice in our multilayers
is in a strong 3D regime.

\vspace{1cm}
\begin{figure}[h]
\begin{center}
\leavevmode \epsfxsize=7cm \epsfysize=5cm \epsffile{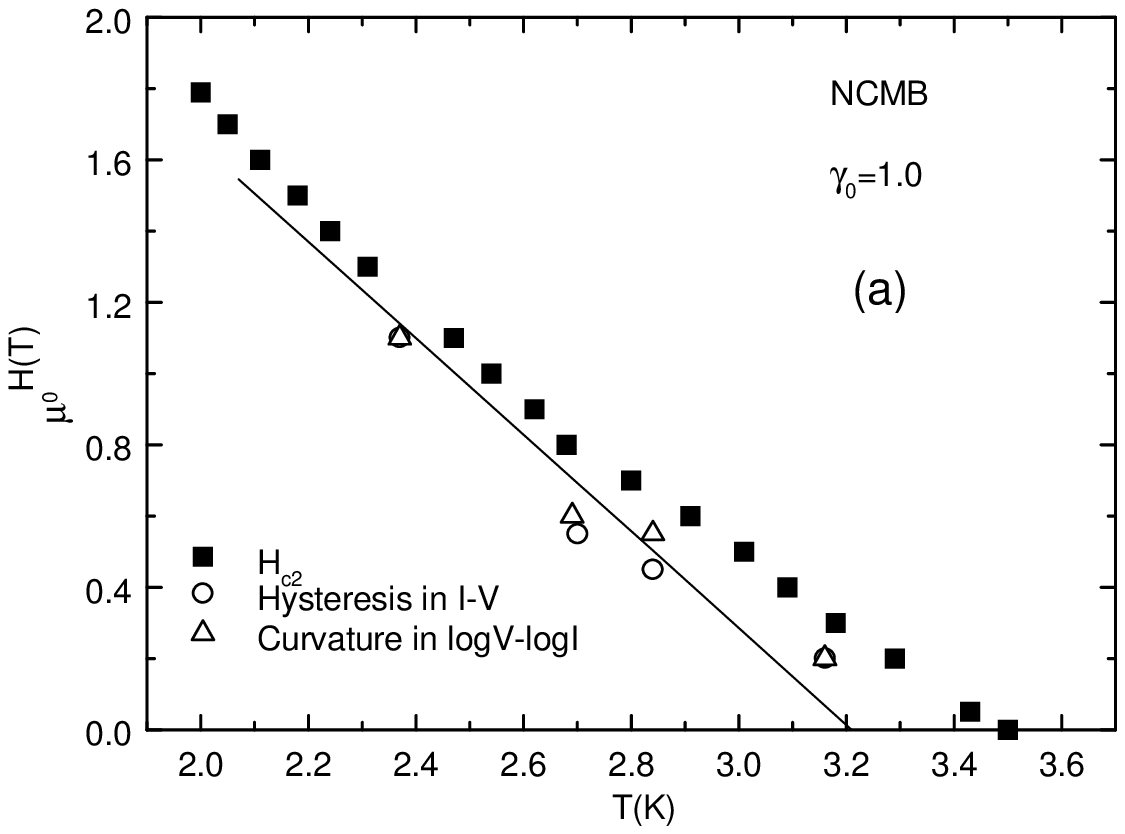}
\end{center}
\end{figure}

\begin{figure}[h]
\begin{center}
\leavevmode \epsfxsize=7cm \epsfysize=5cm \epsffile{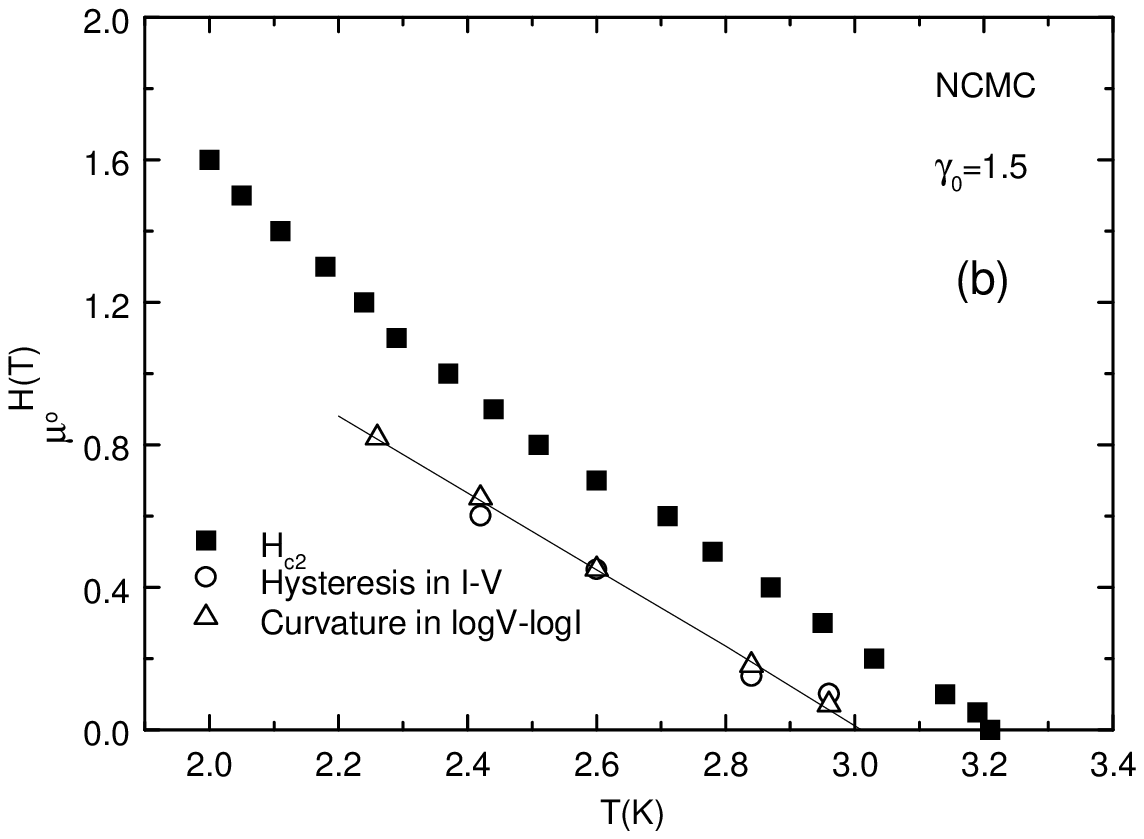}
\end{center}
\end{figure}

\begin{figure}[h]
\begin{center}
\leavevmode \epsfxsize=7cm \epsfysize=5cm \epsffile{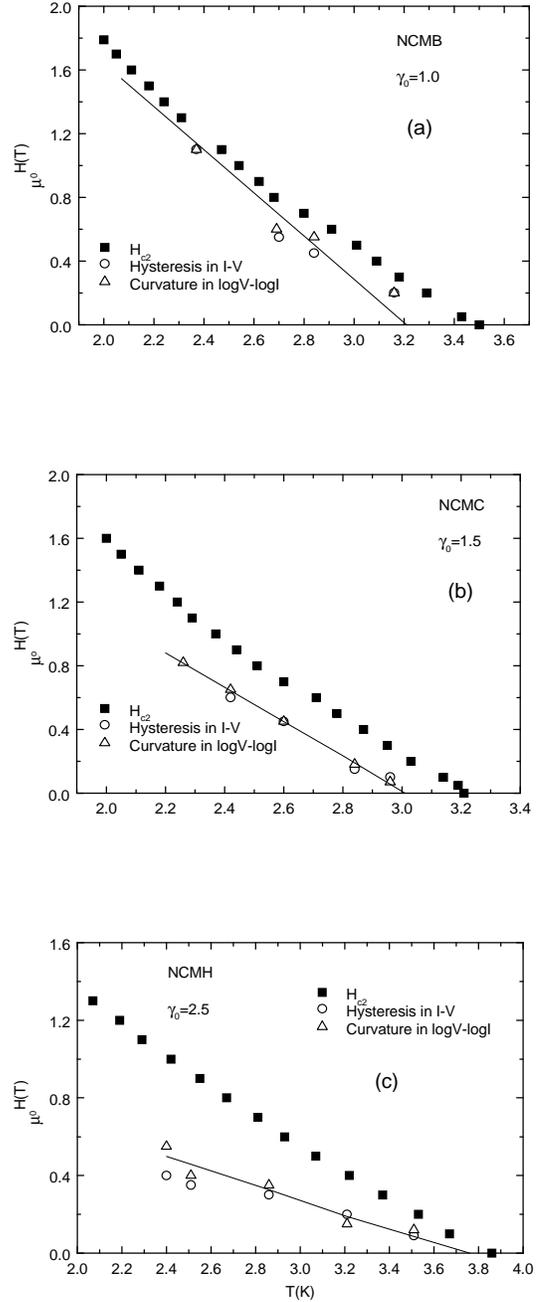}
\end{center}
\caption{\label{fig4}{$H-T$ phase diagram for the sample (a) NCMB,
(b) NCMC and (c) NCMH. Squares correspond to $H_{c2 \perp}$,
circles correspond to the disappearence of hysteresis in the $I-V$
curves, up triangles correspond to the change of the curvature in
the logV-logI curves. The solid lines are calculated according to
the equation 3.}}
\end{figure}

\vspace{1.5cm}
\begin{figure}[h]
\begin{center}
\leavevmode \epsfxsize=7.5cm \epsfysize=5.2cm \epsffile{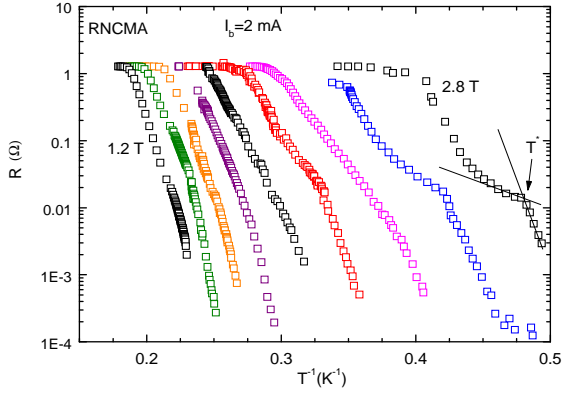}
\end{center}
\caption{\label{fig5}{Arrhenius plot of the transition curves for
the sample RNCMA in perpendicular magnetic fields. The increasing
magnetic field velues are: 1.2; 1.4; 1.6; 1.8; 2.0; 2.2; 2.4; 2.6;
2.8 T. The numbers indicate some values of the magnetic field. The
curves have been recorded using a bias current $I_b$= 2 mA.}}
\end{figure}

\vspace{1.2cm}
\begin{figure}[h]
\begin{center}
\leavevmode \epsfxsize=7.5cm \epsfysize=5.2cm \epsffile{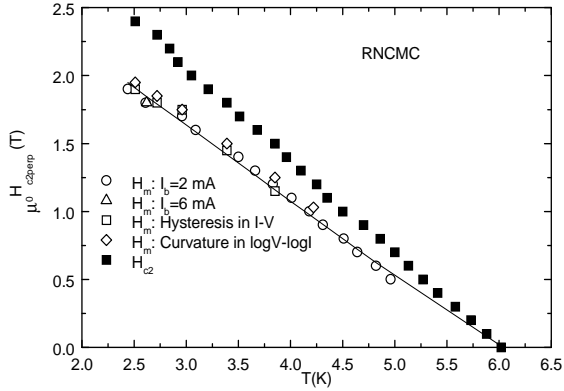}
\end{center}
\caption{\label{fig6}{$H-T$ phase diagram for the sample RNCMC.
Solid squares correspond to $H_{c2 \perp}$, open squares
correspond to the disappearence of hysteresis in the $I-V$ curves,
diamonds correspond to the change of the curvature in the
logV-logI curves, circles (up triangles) correspond to the
$T^{*}(H)$ values extracted from the Arrhenius plot using a bias
current of 2 mA (6 mA). The solid line has been calculated
according to the equation 3.}}
\end{figure}

One of the proposed interpretation of the nature of the IL relies
upon a depinning mechanism in which a crossover from flux creep to
flux flow occurs \cite{Matsu}. In our samples we never obtain
linearity of the $I-V$ curves neither at high voltage where a
uniform flux flow should be present, nor at small currents, where
thermally assisted flux flow (TAFF) \cite{Kes} should take place.
If we fit our data with the relation $V \sim I^{\alpha}$ the
fitting exponent is always very high ($\alpha > 10$). Therefore we
exclude the possibility of flux creep-flux flow crossover in the
presence of a pinning strength distribution which could also be
responsible for the curvature change of the $I-V$ curves when
plotted in double logarithmic scale \cite{Lyk}.

The $I-V$ curves shown in figure \ref{fig2}a and \ref{fig2}b are
very similar to those obtained as a result of numerical
simulations for a superconductor with periodic pinning close to
the matching field \cite{Nori1,Nori2}. Considering that we are
very far from $H_n$ we cannot apply the results of Ref.
\cite{Nori1} to explain our $I-V$ curves. Nevertheless, the main
observed features indicate the presence of a region in the $H-T$
plane where plastic vortex motion probably takes place. Below some
crossover value $H_{pl}$, vortices experience plastic motion which
usually reveals in hysteretical curves \cite{Higg}. Above
$H_{pl}$, $I-V$ curves are smooth, hysteresis vanish and the
vortex motion start to be a flow of vortex liquid \cite{Higg}. The
vortex melting scenario along with the simultaneous presence of
weak and strong pinning channels, can also explain the observed
Arrhenius plots. In fact, immediately below $T_c^{onset}$ (defined
as the temperature where the electrical resistance $R$ is 0.9 of
its value in the normal state), the solidification of the vortices
in the strong pinning channels determines the high values of the
initial slope in the plots. At lower temperatures, the dissipation
in the system is mainly due to the vortices in the weak pinning
regions, and this results in a lower value of the activation
energies. When also these vortices experience the transition from
liquid to solid at $ T=T^*$, the slope in the plots increases
again \cite{Past}.

Melting in the vortex lattice can be induced by thermal
fluctuations \cite{Sudbo}. The melting temperature at which the
vortex lattice goes from a solid phase to a liquid one, can be
obtained by using the 3D thermal melting criterion \cite{Sudbo}

\begin{equation}
c_L^4 \approx
{3G_i \over \pi^2}
{h \over (1-h)^3}
{t_m^2 \over (1-t_m)}
\label{eq:sudbo}
\end{equation}

\noindent where $c_L$ is the Lindemann number, $T_c$ is the
superconducting transition temperature (defined in our case at the
point where the electrical resistance $R$ of the sample becomes
less than $10^{-4}$ $\Omega$), $t_m=T_m /T_c$ is the reduced
melting temperature, and $G_i$ is the Ginzburg number  which
determines the contribution of the thermal fluctuations to the
vortex melting given by $G_i=(1/2)(2 \pi \mu_0 k_B T_c
\lambda_{\parallel}(0) \gamma_0 / \Phi_0^2 \xi_{\parallel}(0))$,
where $\lambda_{\parallel}$ is the in-plane penetration depth
which, for all our samples, has been assumed to be equal to 1500
\AA\ \cite{Lucia}. Melting usually occurs when $c_L \approx 0.1
\div 0.3$ \cite{Blatt2}. When we try to fit the IL observed in our
samples with Eq.2 we get $c_L \sim 10^{-4}$. This extremely low
value for the Lindemann number makes unreasonable to consider the
IL as due to 3D thermal melting.

However, as first pointed out by Blatter and Ivlev \cite{Blatt2}
in moderately anisotropic superconductors at low temperatures one
could not exclude the contribution of quantum fluctuations to the
melting. In this case the total fluctuation displacement of vortex
line is $<u>^2=<u>^2_{th} + <u>^2_q$, where $\sqrt{<u>^2_{th}}$ is
the average displacement due to thermal fluctuations and
$\sqrt{<u>^2_q}$ is the average displacement due to quantum
fluctuations. $<u>^2_{th}$ diminishes with the temperature, while
$<u>^2_q$ does not depend on the temperature and at low values of
T one could expect $<u>^2_{th} \leq <u>^2_q$. The amplitude of
$<u>^2_q$ depends on the ratio $Q^*/\sqrt{G_i}$ where $Q^*=e^2
\rho_N /\hbar s$, with $\hbar$ the Planck constant and $e$ the
elementary charge. If $Q^*/\sqrt{G_i} \gg 1$, the contribution of
quantum fluctuations is crucial. For the samples discussed here we
always get $Q^*/\sqrt{G_i} > 30$ which justifies the possibility
of an important contribution coming from quantum fluctuations
\cite{Volok}. In this case the melting line is given by
\cite{Blatt3}
\begin{equation}
h_m={4 \Theta^2 \over \left[1+\left(1+ 4Q \Theta {1 \over t}
\right)^{1/2}\right]^2} \label{eq:qm}
\end{equation}

\vspace{.3cm} \noindent where $h_m=H_m/H_{c2}$, $t=T/T_c$ is the
reduced temperature, $\Theta=\pi c_L^2(t^{-1}-1) / \sqrt{G_i}$,
$Q=Q^* \Omega \tau /\pi \sqrt{G_i}$ with $\Omega$ a cut-off
frequency generally of the order of the Debye frequency and $\tau$
an effective electronic relaxation time $(\hbar /k_B T_c \approx
\tau)$ \cite{Blatt3}. As we already showed before \cite{Atta}, the
values of $\Omega$ and $\tau$ in Nb/CuMn are in the range $(2 \div
3) \times 10^{13}s^{-1}$ and $(1 \div 5) \times 10^{-13} s$
respectively. Therefore, for all the samples studied we have
assumed $ \Omega=3 \times 10^{13} s^{-1}$ and $ \tau=5 \times
10^{-13}$ s. In this way we reduce the number of free fit
parameters in Eq.3 to only one, namely the Lindemann number $c_L$.

The solid line in Fig.6 has been calculated according to Eq.3
using for the Lindemann number a value of $ c_L$=0.23. Also the
solid lines in the Figures \ref{fig4}a, \ref{fig4}b and
\ref{fig4}c have been calculated according to Eq.3. For the sample
NCMC, Fig.4b, we obtain good agreement with the experimental data
for $c_L=0.19$, while for the sample NCMH, Fig.4c, the solid line
is obtained taking $c_L=0.09$. The agreement between experimental
data and theoretical curves is very good for all the samples
studied. The $c_L$ values, as shown in Table I, become smaller
with increasing anisotropy reaching in the case of sample NCMH a
value slightly below 0.1 \cite{Blatt2}. When increasing the
anisotropy of the system, the coupling between adjacent
superconducting layers reduces and vortex lines become more soft.
The influence of the thermal fluctuations on the vortex dynamics
is strongly dependent on this coupling. Qualitatively, therefore,
the reduction of the $c_L$ values with increasing anisotropy could
be related to the change in the topology of the vortex system.

On the other hand, the softening of the vortex system could also
determine a situation in which thermal fluctuations are able to
cause tilt deformations. Anyway, if we try to fit the experimental
points in figures \ref{fig4} and \ref{fig6} using the 2D pure
thermal melting curve \cite{Kor}
\begin{equation}
{\alpha d \over \kappa^2}
{H_{c2}(T) \over T(1.25-0.25t)}
(1-0.58h_m-0.29h_m^2)
(1-h_m)^2=1\label{eq:kor}
\end{equation}

\noindent we do not obtain any agreement with the data. Here
$\alpha=A \Phi_0 (1.07)^2 / 32 \pi \mu_0 k_B$,
$\kappa=\lambda_{\parallel} / \xi_{\parallel}$ and $d$ is the
thickness of the sample. $A$ is a renormalization factor of the
shear modulus $c_{66}$ due to non linear lattice vibrations and
vortex lattice defects and is $A \approx 0.64$ \cite{De}.

\vspace{1.5cm}
\begin{figure}[h]
\begin{center}
\leavevmode \epsfxsize=7cm \epsfysize=5cm \epsffile{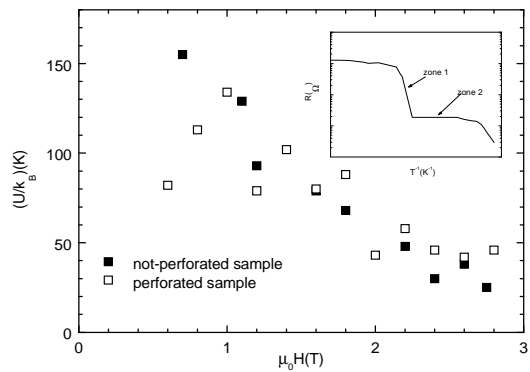}
\end{center}
\caption{\label{fig7}{Values of the activation energy in the zone
2 for a not-perforated sample (solid squares) and values of the
activation energy in the zone 1 for a perforated sample (open
squares). Insert: schematic of a typical Arrhenius plot observed
in a perforated and a not-perforated Nb/CuMn multilayer together
with the identification of the zones 1 and 2.}}
\end{figure}

We want to point out that the quantum melting theory has been
succesfully applied to describe the vortex behavior also in
not-perforated Nb/CuMn multilayers \cite{Atta}. In that case the
melting line was determined by analyzing in Arrhenius fashion the
measured $R(T)$ curves in perpendicular magnetic fields. The
shapes of the Arrhenius plots, see for example figure \ref{fig1}
in ref. \cite{Atta}, were very similar to those observed in the
case of perforated samples, suggesting the presence of two types
of pinning centers also in the case of non perforated samples. In
not perforated samples, edge pinning could be relevant and
obviuosly stronger than intrinsic pinning \cite{Bean}. Therefore
one could interpret the shape of the Arrhenius plots in not
perforated Nb/CuMn multilayers as due to vortex transition from
liquid to solid first of the vortices at the edges and then, at
lower temperatures, of the vortices intrinsically pinned in the
inner part of the samples. If this interpretation is correct, the
slopes of the Arrhenius plots measured in not perforated samples
in the zone 2 (see insert in figure \ref{fig7}) should be very
close to the slope measured in the zone 1 in perforated samples.
In fact, in both cases, this slope should be related to the
activation energy of vortices intrinsically pinned inside the
system. In figure \ref{fig7} the solid points refer to the values
of the activation energy in a typical not-perforated sample
measured in zone 2, while the open symbols refer to the values of
the activation energy in a perforated sample (RNCMA) measured in
zone 1. The two samples have been chosen to have similar critical
temperatures. Also the $R(T,H)$ curves have been takin using a
similar value for the bias current density $J_b$. The quite nice
agreement between the two sets of data supports our idea that, in
both cases, they are a measure of the intrinsic pinning in the
material.

The presence of the antidot array in the multilayers makes more
easily measurable the vortex melting. In fact, in the case of
perforated samples one is able to detect the change in the slope
of the Arrhenius plots using low values of the bias current ($
\sim 100 \mu $A) while for not perforated samples bias currents of
$ \sim 1 mA$ are needed to observe the same effect (the IL in a
superconductor does not depend on the value of the bias current).
This is consistent with the idea that in antidotted samples the
melting takes place in the zones with weaker pinning when compared
to the case of not perforated samples in which the measured vortex
phase transition takes place in the zones of intrinsic pinning. As
a consequence of this also the hysteresis is much easier
detectable in antidotted samples, in regions of the $I-V$
characteristics not to close to the $H_{c2}$(T) curve.

\begin{figure}[h]
\begin{center}
\leavevmode \epsfxsize=7.5cm \epsfysize=5.5cm \epsffile{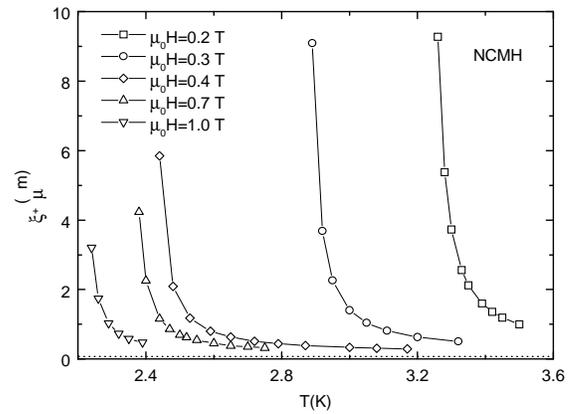}
\end{center}
\caption{\label{fig8}{Dependence of the correlation length
$\xi_{+}$ versus the temperature at different magnetic fields for
the sample NCMH.}}
\end{figure}

The influence of the regular array of antidots on the vortex
properties is also confirmed by the behavior in our samples of the
vortex correlation length in the liquid phase $\xi_{+}$, defined
as \cite{Nelson}

\begin{equation}
\xi_{+} \approx \xi_{+0} \exp \left \{b\left( {T_m \over T-T_m}
\right)^{\nu} \right \} \label{eq:csi}
\end{equation}

\noindent where $b$ is a constant of the order of the unity,
$\nu$=0.36963 and $\xi_{+0}$, being the smallest characteristic
length scale in the liquid, is of the order of $a_0$, the vortex
lattice parameter. According to the melting theory the shear
viscosity $\eta(T)$ of the vortex liquid starts to grow up
approacching the liquid-solid line from above, when $\eta \sim
\xi_{+}^2(T)$ \cite{Nelson}. The curves of $\xi_{+}$ versus the
temperature for different applied magnetic fields are reported in
figure \ref{fig8} for the sample NCMH. All the curves start to
diverge when the value of $\xi_{+}$ becomes comparable to the
average distance among the antidots, i.e. when $\xi_{+} \approx
d=$1 $\mu m$. Similar results have been obtained for all the
samples investigated. This is exactly what we expect, if we look
at the investigated system as a vortex ensemble constrained in the
narrow channels among the lines of the antidots. In this case,
infact, the melting transition at $H=H_m$ is going to be observed
when the correlation length of the liquid $\xi_{+}$ reaches the
value of the width of the channels. The results shown in figure
\ref{fig8} clearly indicate the influence of the antidots lattice
on the vortex dynamics in our samples.

\vspace{1.5cm}

In conclusion, we have studied transport properties of
superconducting (Nb)-spin glass (CuMn) multilayer with a regular
array of antidots by measuring $I-V$ curves in perpendicular
magnetic fields. The measurements have been performed far above
the matching conditions. The dynamic phase diagram has been drawn
out from the analysis of these measurements. Two regions
corresponding to plastic flux flow motion and to the motion of the
vortex liquid have been distinguished. Melting occurs mostly due
to quantum fluctuations and the presence of the antidots renders
more easier to detect the melting due to the weaker pinning in the
zones with higher local current density.


\begin{references}
\bibitem{Camp}  A.M. Campbell and J.E. Evetts, Adv. Phys.
{\bf 21}, 199 (1972).

\bibitem{Blatt1} G. Blatter, M.V. Feigelman, V.B. Geshkenbein, A.I. Larkin,
and V.M. Vinokur, Rev. Mod. Phys. {\bf 66}, 1125 (1994).

\bibitem{Nb3Sn} K. Tachikawa, Y. Kuroda, H. Tomori, and M. Ueda, IEEE Trans. Appl. Superconduc. {\bf 7}, 1355 (1997).

\bibitem{Shu}  J.J. Martin, M. Vélez, J. Noguès, and I.K. Schuller, Phys. Rev. Lett. {\bf 79}, 1929 (1997).

\bibitem{Civ} L. Civale, A. Marwick, T.K. Worthington, M.A. Kirk, J.R. Thompson, L. Krusin-Elbaum, Y. Sun, J.R. Clem, and F. Holtzberg, Phys. Rev. Lett. {\bf 67}, 648 (1991).

\bibitem{Pru}  A. Pruijmboom, P.H. Kes, E. van der Drift, and S. Radelaar,
Phys. Rev. Lett. {\bf 60}, 1430 (1988).

\bibitem{Crabtree} W.K. Kwok, J.A. Fendrich, V.M. Vinokur, A.E. Koshelev, and G.W. Crabtree,
Phys. Rev. Lett. {\bf 76}, 4596 (1996).

\bibitem{Bae}  M. Baert, V.V. Metlushko, R. Jonckheere, V.V. Moshchalkov, and Y. Bruynseraede,  Phys. Rev. Lett. {\bf 74}, 3269 (1995).

\bibitem{Ket}  D.J. Morgan and J.B. Ketterson, Phys. Rev. Lett. {\bf 80}, 3614 (1998).

\bibitem{Word} A. Castellanos, R. W\"ordenweber, G. Ockenfuss, A.v.d. Hart, and K. Keck, Appl. Phys. Lett. {\bf 71}, 962 (1997).

\bibitem{Fio} A.T. Fiory, A.F. Hebard, S. Somekh, Appl. Phys. Lett. {\bf 32}, 73 (1978).

\bibitem{Lykov} A.N. Lykov, Solid State Commun. {\bf 86}, 531 (1993).

\bibitem{Nori1}  C. Reichhardt, C.J. Olson, and F. Nori, Phys. Rev. Lett. {\bf 78}, 2648 (1997).

\bibitem{Nori2}  C. Reichhardt, C.J. Olson, and F. Nori, Phys. Rev. B {\bf 58}, 6534 (1998).

\bibitem{Mart}  M.H. Theunissen, E. Van der Drift, and P.H. Kes, Phys. Rev. Lett. {\bf 77}, 159 (1996).

\bibitem{Past}  H. Pastoriza and P.H. Kes, Phys. Rev. Lett. {\bf 75}, 3525 (1995).

\bibitem{Philmag} C. Attanasio, T. Di Luccio, L.V. Mercaldo, S.L. Prischepa, R. Russo, M. Salvato, L. Maritato, and S. Barbanera, accepted for publication in Philos. Mag. B (2000).

\bibitem{Multasmodel} W.R. White, A. Kapitulnik, and M.R. Beasley, Phys. Rev. Lett. {\bf 66}, 2826 (1991).

\bibitem{Blatt2}  G. Blatter and B.I. Ivlev, Phys. Rev. Lett. {\bf 70}, 2621 (1993).

\bibitem{Maritato}  L.V. Mercaldo, C. Attanasio, C. Coccorese, L. Maritato, S.L. Prischepa, and M. Salvato, Phys. Rev. B {\bf 53}, 14040 (1996).

\bibitem{Tin}  M. Tinkham, {\sl Introduction to Superconductivity}, (Mc Graw-Hill, New York, 1996).

\bibitem{Berg}  P. Berghuis, A.L.F. van der Slot, and P.H. Kes,  Phys. Rev. Lett. {\bf 65}, 2583 (1990).

\bibitem{Bae1}  M. Baert, V.V. Metlushko, R. Jonckheere, V.V. Moshchalkov, and Y. Bruynseraede, Europhys. Lett. {\bf 29}, 157 (1995).

\bibitem{Vin} V.H. Vinokur, P.H. Kes, and A.E. Koshelev, Physica C {\bf 169}, 29 (1990).

\bibitem{Glaz} L.I. Glazman and A.E. Koshelev, Phys. Rev. B {\bf 43}, 2835 (1991).

\bibitem{Matsu} T. Matsushita and T. Kiss, Physica C {\bf 315}, 12 (1999).

\bibitem{Kes}  P.H. Kes, J. Aarts, J. van den Berg, C.J. van der Beek, and J.A. Mydosh, Superc. Sci. Technol. {\bf 1}, 242 (1989).

\bibitem{Lyk}  A.N. Lykov, C. Attanasio, L. Maritato, and S.L. Prischepa, Superc. Sci. Technol. {\bf 10}, 119 (1997).

\bibitem{Higg}  M.J. Higgings and S. Bhattacharya, Physica C {\bf 257}, 232 (1996).

\bibitem{Sudbo}  A. Houghton, R.A. Pelcovits, and A. Sudb\o, Phys. Rev. B {\bf 40}, 6763 (1989).

\bibitem{Lucia} L.V. Mercaldo, S.M. Anlage, and L. Maritato, Phys. Rev. B {\bf 59}, 4455 (1999).

\bibitem{Volok}  G. Blatter, B.I. Ivlev, Yu. Kagan, M.H. Theunissen, Y. Volokitin, and P.H. Kes, Phys. Rev. B {\bf 50}, 13013 (1994).

\bibitem{Blatt3}  G. Blatter and B.I. Ivlev, J. Low Temp. Phys. {\bf 95}, 365 (1994).

\bibitem{Atta}  C. Attanasio, C. Coccorese, L. Maritato, S.L. Prischepa, M. Salvato, B. Engel, and C.M. Falco, Phys. Rev. B {\bf 53}, 1087 (1996).

\bibitem{Kor} P. Koorevaar, Ph.D. Thesis, Leiden University (1994).

\bibitem{De} S.W. de Leeuw and J.W. Perram, Physica A {\bf 113}, 546 (1982).

\bibitem{Bean} C.P. Bean and J.D. Livingston, Phys. Rev. Lett. {\bf 12}, 14 (1964).

\bibitem{Nelson} D.R. Nelson and B.I. Halperin, Phys. Rev. B {\bf 19}, 2457 (1979).



\end{references}
\end{document}